\documentclass{iopart}
\usepackage{iopams}
\usepackage{graphicx}
\graphicspath{{./images/}}
\usepackage{xcolor}

\usepackage{graphicx}
\usepackage{enumitem}
\usepackage{ulem} %added by SD. Allows strikeout command \sout{}  
\usepackage{geometry}

\expandafter\let\csname equation*\endcsname\relax
\expandafter\let\csname endequation*\endcsname\relax
 \usepackage{amsmath}

\usepackage{xifthen}% provides \isempty test
\newcommand{\vq}[2][]{                % vq = Vector of Quadratures
  \ifthenelse{\isempty{#1}}           %
    { \hat{\pmb{#2}} }             % if #2 is empty
    { \hat{\pmb{#2}}_\mathrm{#1} } % if #2 is not empty
}
\newcommand{\tq}[2][]{                % tq = matrix of Tranformation of Quadratures
  \ifthenelse{\isempty{#1}}           %
    { \mathbb{#2} }                   % if #2 is empty
    { \mathbb{#2}^\mathrm{#1} }       % if #2 is not empty
}
\newcommand{\vs}[2][]{                % vq = Vector of Scalars
  \ifthenelse{\isempty{#1}}           %
    { \mathbf{#2} }                   % if #2 is empty
    { \mathbf{#2}^\mathrm{#1} }       % if #2 is not empty
}

\begin{document}

\title{Quantum noise cancellation in asymmetric speed meters with balanced homodyne readout}

\author{
T~Zhang$^1$
E~Knyazev$^{3}$,
S~Steinlechner$^{1,4}$,
F~Ya~Khalili$^{3,5}$,
B~W~Barr$^1$,
A~S~Bell$^1$,
P~Dupej$^1$,
C~Gr\"af$^1$,
J~Callaghan$^1$
J~S~Hennig$^1$,
E~A~Houston$^1$,
S~H~Huttner$^1$,
S~S~Leavey$^1$,
D~Pascucci$^1$,
B~Sorazu$^1$,
A~Spencer$^1$,
J~Wright$^1$,
K~A~Strain$^1$,
S~L~Danilishin$^{1,2}$
and 
S~Hild$^1$
}
\ead{stefan.danilishin@ligo.org}
\vskip 1mm
\address{$^1$ School of Physics and Astronomy, The University of Glasgow, Glasgow, G12\,8QQ, UK\\
$^{2}$\,Institut f\"ur Theoretische Physik, Leibniz Universit\"at Hannover and Max-Planck-Institut f\"ur Gravitationsphysik (Albert-Einstein-Institut), Callinstra\ss e 38, D-30167 Hannover, Germany\\
$^{3}$\,Faculty of Physics, M.V. Lomonosov Moscow State University, 119991 Moscow, Russia\\
$^{4}$  Institut f\"ur Laserphysik und Zentrum f\"ur Optische Quantentechnologien der Universit\"at Hamburg, Luruper Chaussee 149, 22761 Hamburg, Germany\\
$^{5}$ Russian Quantum Center, 143025 Skolkovo, Russia}
\begin{abstract}
Sagnac  speed meter (SSM) topology is known  as an alternative technique to reduce quantum  back-action in  gravitational-wave interferometers. However, any potential imbalance of the main beamsplitter was shown to reduce the quantum noise superiority of speed meter at low frequencies, caused due to increased laser noise coupling to the detection port. In this paper, we show that implementing balanced homodyne readout scheme and for a particular choice of the local oscillator (LO) delivery port, the excess laser noise contribution to quantum noise limited sensitivity (QNLS) is partly compensated and the speed meter sensitivity can outperform state-of-the-art position meters. This can be achieved by picking the local oscillator from interferometer reflection (\textit{co-moving} LO) or the main beamsplitter anti-reflective coating surface (BSAR LO). We also show that this relaxes the relative intensity noise (RIN) requirement of the input laser. For example, for a beam splitter imbalance  of $0.1 \%$ in Glasgow speed meter proof of concept experiment, the RIN requirement at frequency of 100Hz decreases from $4\times 10^{-10}/\sqrt{\rm Hz}$ to $4\times 10^{-7}/\sqrt{\rm Hz}$, moving the RIN requirement from a not practical achievable value to one which is routinely achieved with moderate effort.
\end{abstract}

\section{Introduction}
\label{sec:intro}
In 2015, we have stepped into the new era of gravitational-wave astronomy with the first direct detection of gravitational waves (GW) from the colliding binary black hole (BBH) system by the two Advanced LIGO interferometers \cite{FirstDetection}. The two exciting years of discoveries that followed have given us four more BBH merger events \cite{SecondDetection,PhysRevLett.118.221101,2017_PRL.119.141101_LVC,2017ApJ.851L35A_LVC_Detection} and one collision of neutron stars \cite{2017_PRL.119.1101A_LVC_BNS_Discovery} observed also in the electromagnetic spectrum \cite{2017_ApJ.848L.12A_LVC}.

Those serendipitous discoveries, apart from a great deal of fascinating new science hitherto unavailable to men, have identified the need to improve the sensitivity of the existing detectors in the low frequencies ($<30$~Hz). The low-frequency end of the GW detector's noise is currently hampering the access to the GW signals from the population of massive black holes (with masses $>30M_\odot$, where $M_\odot$ is the solar mass) revealed by the recent detections, as well as covering GWs from the early inspiral stage of binary neutron stars evolution crucial for issuing an early warning to EM telescopes.

Modern gravitational-wave detectors are limited in sensitivity over much of their detection frequency band by quantum noise \cite{Harry10,Abbott2018} that stems from the fundamental quantum-mechanical fluctuations of laser light phase and amplitude. In particular, amplitude fluctuations create random back-action force mimicking the action of GWs that has the largest impact at low frequencies where noise amplitude rises as $f^{-2}$, where $f$ is the GW frequency. 

Speed-meter interferometers were first proposed by Braginsky and Khalili \cite{Braginsky90} as a way to suppress quantum back-action noise in bar GW detectors. Later, this concept was generalised to laser GW interferometers \cite{brag_2000}. The back-action noise reduction in speed meters stems from the quantum non-demolition (QND) nature of test mass' velocity \cite{QNDreview} as quantum observable, as opposed to the displacement measured by Michelson interferometers. This advantage of speed meters over position meters at low frequencies inspired the development of several different topologies of speed meters \cite{Purdue02,Chen2002,Danilishin04,Wang13,2017_CQG.34.2.024001_Huttner,KNYAZEV2017}. 

One of these configurations, the zero-area Sagnac interferometer, was first identified by Chen\cite{Chen2002} as a QND speed-meter. It performs a relative speed measurement of the test masses in the arms by letting the two counter propagating light beam visit both arms sequentially in the opposite directions. Thus each beam, upon returning to the main beam-splitter, carries in its phase the information about the sum of the two arms displacements separated by a time delay $\tau$ equal to the arm cavity ringdown time. After recombination at the beam-splitter, the light at the readout port of the interferometer has the phase proportional to the mean relative velocity of the interferometer arm length change, hence it performs a QND measurement of speed.

In an ideal case of perfectly symmetric beam splitter, Sagnac interferometer is always operating at the dark fringe at DC independent of the tuning of the arms and only the signal sidebands proportional to the relative differential velocity of the interferometer's arms propagate to the readout port. This robustness of Sagnac topology to optical path variations compared to Michelson one was deemed as an advantage warranting its implementation as GW detectors \cite{Sun0}. However it was recognised later on that any deviation from the 50:50 ratio in its main beamsplitter would pose a limit to the achievable sensitivity of a Sagnac interferometer due to coupling of laser port fluctuations to the readout port \cite{Beyersdorf, SagnacImperfections}. 

\begin{figure}[htbp]
	\centering
	\includegraphics[width=\textwidth]{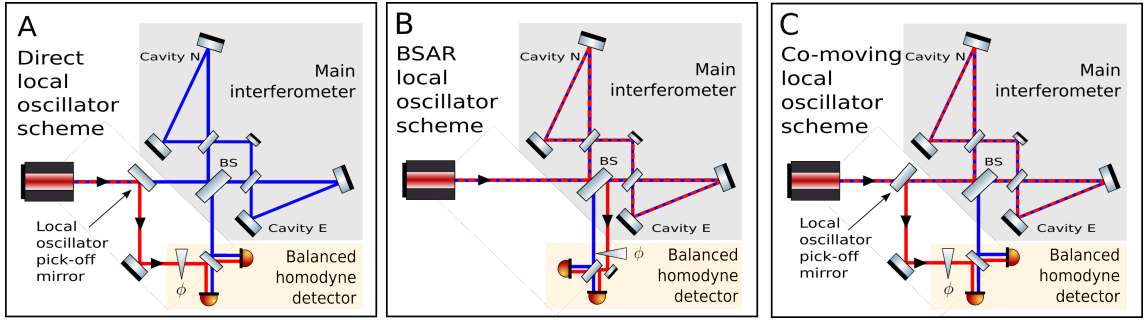}
	\caption{Topologies considered for the speedmeter with balanced homodyne detector(BHD). Blue lines represent the path of the laser light through the interferometer,  red dashed lines represent the shared path of the local oscillator and interferometer light, and the red solid line represent the local oscillator after its path diverges from the interferometer light. $A$ shows the case where the local oscillator is derived by tapping off a small fraction of the input beam and guiding it to the output port. $B$ shows the case where the local oscillator is derived by tapping off the intercavity light at the central beamsplitter's anti-reflective coating. $C$ shows the case where the light used as the local oscillator will have passed through the whole interferometer and encountered the same delay and dispersion as well as the same optomechanical interaction as the signal beam.}	
	\label{fig:ssm_cavs_bhd}
\end{figure}

Blending speedmeter topologies with certain readout methods has been shown to provide a partial reduction in the coupling of laser noise fluctuation to the gravitational wave readout signal \cite{Sun:97:01, Sun:97:02, Beyersdorf}. In this paper we take inspiration from that, and analytically investigate the potential cancellation of quantum noise in asymmetric Sagnac speedmeters that use balanced homodyne detectors. In extending this analysis to the Glasgow Sagnac speed meter, we investigate the potential additional cancellation of laser technical noise. The Glasgow Sagnac speed meter will use a balanced homodyne readout  and we show three LO delivery options as per Fig.~\ref{fig:ssm_cavs_bhd}. Here we examine the quantum and classical noise reduction when using a balanced homodyne detector local oscillator taken from the interferometer bright port versus the noisier option of using the laser light which has not been through the interferometer. Given this noise cancellation, we can allow for deviation from a 50:50 main beamsplitter ratio and thus resolve the main problem that has been identified with Saganac interferometers.

In Sec.~\ref{sec:theory}, we conduct an analytical treatment of quantum noise of an asymmetric  Sagnac speed meter interferometer, and show how balanced homodyne readout can help to suppress quantum noise with the proper choice of the LO. In Sec.~\ref{sec:glas} we show the analysis on the relaxed requirement of relative laser intensity noise base on simulation software FINESSE\cite{finesse}.

\section{Quantum Noise of an Imperfect Speedmeter IFO}
\label{sec:theory}
\subsection{Two-photon formalism.}
\begin{figure}[htbp]
	\centering
	\includegraphics[width=0.4\textwidth]{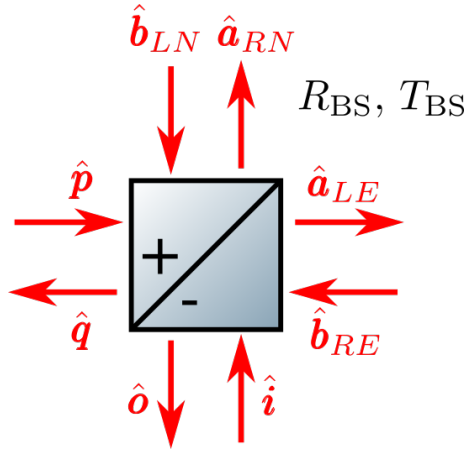}
	\caption{Schematic of the input and output fields around the main beam splitter.}
	\label{fig:bs}
\end{figure}

In this section, we use the two-photon formalism of quantum optics \cite{CavesTwoPhotonTheory,85a2CaSch}. It describes, locally, an arbitrary quasi-monochromatic modulated electromagnetic wave with strain $ \hat{E}(t) = \mathcal{E}_0\left[(A_c+\hat{a}_c(t))\cos\omega_p t+(A_s+\hat{a}_s(t))\sin\omega_pt\right]$ in terms of 2-dimensional vectors of quadrature amplitudes $\pmb{A}+\hat{\pmb{a}}$, where $\pmb{A} = \{A_c,\,A_s\}^{\rm T}$ stands for DC mean amplitudes vector and $\hat{\pmb{a}} = \{\hat a_c,\,\hat a_s\}^{\rm T}$ stands for zero-mean non-stationary variations and fluctuations of light (superscript $^{\rm T}$ denotes transpose of the matrix or vector). Here normalisation constant $\mathcal{E}_0 = \sqrt{\frac{4\pi\hbar\omega_p}{\mathcal{A}c}}$, $\mathcal{A}$ is effective cross section of the beam, $c$ the speed of light, and $\omega_p$ is the carrier light frequency. It is usually more convenient to work in the frequency domain:
\begin{equation}
  \hat{a}_{c,s}(t) = \int_{-\infty}^\infty \frac{d\Omega}{2\pi} \hat{a}_{c,s}(\Omega)e^{-i\Omega t}\,,
\end{equation} 
where we define quadratures spectra at the modulation sidebands off-set frequency $\Omega = \omega-\omega_p$.

In order to understand how the fluctuations, entering the pumping port of the interferometer influence all three variants of LO choice, we need to analyse the input-output relations of the asymmetric interferometer with an emphasis on the transfer functions of the pump sideband fields to both, the readout port, and to the LO. Hereinafter we attain the result.

\subsection{Input-output relations of the asymmetric Sagnac interferometer.}

We consider a Sagnac interferometer with non-unity ratio $R_{\rm BS}/T_{\rm BS}\neq1$ depicted in Fig.~\ref{fig:bs}, with $R_{\rm BS}$ and $T_{\rm BS}$ representing the power reflectivity and transmissivity of the main beam splitter. The three LO choices that we investigate here require the knowledge of the following 3 output fields,
\begin{enumerate}[label=(\roman*)]
\item Readout port output field $\pmb{\hat{o}}$ (for all three variants)
\item Part $\pmb{\hat{b}}^{\rm RE}$ of the output field $\pmb{\hat{o}}$ contributed by the clockwise propagating light beam that gives the LO field upon reflection off the main beam splitter anti-reflecting coating (variant Fig.~\ref{fig:ssm_cavs_bhd}B)
\item Return field $\pmb{\hat{q}}$ at the pumping port (for the \textit{co-moving LO} choice of Fig.~\ref{fig:ssm_cavs_bhd}C) 
\end{enumerate}
Expressed in terms of the dark port (DP) input field, $\pmb{\hat{i}}$ and bright port (BP) input field $\pmb{\hat{p}}$ and signal displacements. Following the \cite{SagnacImperfections}, those can be written as:  
\begin{eqnarray}
\pmb{\hat o} &= \mathbb{T}_i\, \pmb{\hat i} +\mathbb{T}_p\, \pmb{\hat p}+\pmb{t}_{\rm d}x_d+\pmb{t}_{\rm c}x_c\,,\label{eq:o}\\
\pmb{\hat q} &= \mathbb{R}_i\, \pmb{\hat i} +\mathbb{R}_p\, \pmb{\hat p}+\pmb{q}_{\rm d}x_d+\pmb{q}_{\rm c}x_c\,,\label{eq:q}\\
\pmb{\hat b}^{\rm RE} &= \mathbb{T}^{\rm RE}_i\, \pmb{\hat i} +\mathbb{T}^{\rm RE}_p\, \pmb{\hat p}+\pmb{t}^{\rm RE}_{\rm d}x_d+\pmb{t}^{\rm RE}_{\rm c}x_c\,,
\end{eqnarray}
where $x_c = x_n+x_e$ and  $x_d = x_n-x_e$ stand for the two mechanical modes of the Sagnac interferometer, namely the \textit{common} and the \textit{differential} arms
elongation modes. The transfer matrices $\mathbb{T}_i $, $\mathbb{T}^{\rm RE}_i $ and $\mathbb{R}_i$ define the coupling of dark port input field $\pmb{\hat i}$ to the corresponding output port. The other three matrices are of more interest to us, \textit{i.e.} the $\mathbb{T}_p $, $\mathbb{T}^{\rm RE}_p$ and $\mathbb{R}_p$, as they describe how laser fluctuations $\pmb{\hat p}$ couple to the corresponding output ports of the interferometer. It is straightforward to show (see \cite{SagnacImperfections} for details) that these transfer matrices, in case of imbalanced beam splitter with $R_{\rm BS}\neq T_{\rm BS}$, follow the well known structure of the tuned optomechanical interferometer transfer matrix (see, \textit{e.g.}, \cite{02a1KiLeMaThVy,Review}):
\begin{eqnarray}\label{eq:TrMat} 
\mathbb{T}_i &= 2\sqrt{R_{\rm BS}T_{\rm BS}} e^{2i\beta_{\rm sag}} \begin{bmatrix}1 & 0\\ -\mathcal{K}_{\rm sym} & 1\end{bmatrix}\,, \\ \mathbb{R}_i &=(R_{\rm BS}-T_{\rm BS}) e^{2i\beta_{\rm sag}} \begin{bmatrix}1 & 0\\ 0 & 1\end{bmatrix}\,,\\ 
\mathbb{T}_p &= (R_{\rm BS}-T_{\rm BS})e^{2i\beta_{\rm sag}}\begin{bmatrix}1 & 0\\ -4\mathcal{K}_{\rm arm} & 1\end{bmatrix}\,,\label{eq:TrMat1} \\ \mathbb{R}_p &=-2\sqrt{R_{\rm BS}T_{\rm BS}}e^{2i\beta_{\rm sag}}\begin{bmatrix}1 & 0\\ -\mathcal{K}_{\rm asym}& 1\end{bmatrix}\,,\label{eq:TrMat2}\\
\mathbb{T}^{\rm RE}_i &= \sqrt{T_{\rm BS}}e^{2i\beta_{\rm sag}}\begin{bmatrix}1 & 0\\ -2R_{\rm BS}\mathcal{K}_{\rm sym} & 1\end{bmatrix}\,, \\ \mathbb{T}^{\rm RE}_p &=\sqrt{R_{\rm BS}}e^{2i\beta_{\rm sag}}\begin{bmatrix}1 & 0\\ 4\mathcal{K}_{\rm arm}-2T_{\rm BS}\mathcal{K}_{\rm sym}& 1\end{bmatrix}\,,
\end{eqnarray}
with diagonal elements describing the purely optical response (with fixed mirrors position), whereas the lower off-diagonal term, featuring the so called optomechanical coupling factor $\mathcal{K}$ first introduced by Kimble \textit{et al.} \cite{02a1KiLeMaThVy}, embraces the details of interaction of mechanical degrees of freedom of the interferometer with the corresponding light field (via radiation pressure). Response of the interferometer to both, differential and common mechanical motion of the mirrors can be written as:
\begin{eqnarray}\label{eq:OMResp}
\pmb{t}_{\rm d} &= -e^{i\beta_{\rm sag}}\dfrac{\sqrt{2\mathcal{K}_{\rm sym}}}{x_{\rm SQL}}\begin{bmatrix}0\\1\end{bmatrix}\,, \\ \pmb{t}_{\rm c} &= ie^{i\beta_{\rm sag}}\dfrac{(R_{\rm BS}-T_{\rm BS})\sqrt{2\mathcal{K}_{\rm asym}}}{x_{\rm SQL}}\begin{bmatrix}0\\1\end{bmatrix}\,,\\ 
\pmb{t}^{\rm RE}_{\rm d} &= e^{i\beta_{\rm sag}}\dfrac{\sqrt{2R_{\rm BS}\mathcal{K}_{\rm sym}}}{x_{\rm SQL}}\begin{bmatrix}0\\1\end{bmatrix}\,, \\ \pmb{t}^{\rm RE}_{\rm c} &= -ie^{i\beta_{\rm sag}}\dfrac{\sqrt{2R_{\rm BS}\mathcal{K}_{\rm asym}}}{x_{\rm SQL}}\begin{bmatrix}0\\1\end{bmatrix}\,,\\ 
\pmb{q}_{\rm d} &=\begin{bmatrix}0\\0\end{bmatrix}\,, \\ \pmb{q}_{\rm c} &= -e^{i\beta_{\rm sag}}\dfrac{2\sqrt{2R_{\rm BS}T_{\rm BS}\mathcal{K}_{\rm asym}}}{x_{\rm SQL}}\begin{bmatrix}0\\1\end{bmatrix}\,,
\end{eqnarray}
where $\beta_{\rm sag} = 2\beta_{\rm arm}+\frac\pi2$ is the Sagnac-specific additional phase shift that signal sidebands at frequency $\Omega$ acquire in the course of propagation through the interferometer. $x_{\rm SQL} = \sqrt{\dfrac{2\hbar}{M\Omega^2}}$ stands for the free mass displacement standard quantum limit (SQL). Symmetric and asymmetric optomechanical coupling factors of imperfect Sagnac interferometer are defined the same way as in \cite{SagnacImperfections}:
\begin{eqnarray}\label{eq:Ksym}
\mathcal{K}_{\rm sym} &= 4\mathcal{K}_{\rm arm}\sin^2\beta_{\rm arm} \simeq \dfrac{8\Theta_{\rm arm}\gamma_{\rm arm}}{(\Omega^2+\gamma_{\rm arm}^2)^2}\,,
\\
\mathcal{K}_{\rm asym} &= 4\mathcal{K}_{\rm arm}\cos^2\beta_{\rm arm} \simeq \dfrac{8\Theta_{\rm arm}\gamma^3_{\rm arm}}{\Omega^2(\Omega^2+\gamma_{\rm arm}^2)^2}\,.
\end{eqnarray}
with $\beta_{\rm arm} = \arctan\frac{\Omega}{\gamma_{\rm arm}}$ the phase shift acquired by a sideband field in one arm cavity. $\gamma_{arm}=\frac{cT_{ITM}}{4L}$ is the half-bandwidth of the arm cavities with length $L$  and input mirror power transmissivity $T_{ITM}$. And $\mathcal{K}_{\rm arm}= \dfrac{2\Theta\gamma_{\rm arm}}{\Omega^2(\gamma^2_{\rm arm}+\Omega^2)}$ is the optomechanical coupling factor of this arm with $\Theta = \frac{4\omega_pP_{\rm arm}}{McL}$ the normalised power, where $P_{\rm arm}$ is the circulating in each arm of an equivalent Michelson, $M$ is  the reduced mass of the dARM mode and $L$ is the length of the arm. Note that $\mathcal{K}_{\rm sym}+\mathcal{K}_{\rm asym} = 4\mathcal{K}_{\rm arm}$, which will be used later.
\begin{figure}[t]
	\centering
	\includegraphics[width=0.8\textwidth]{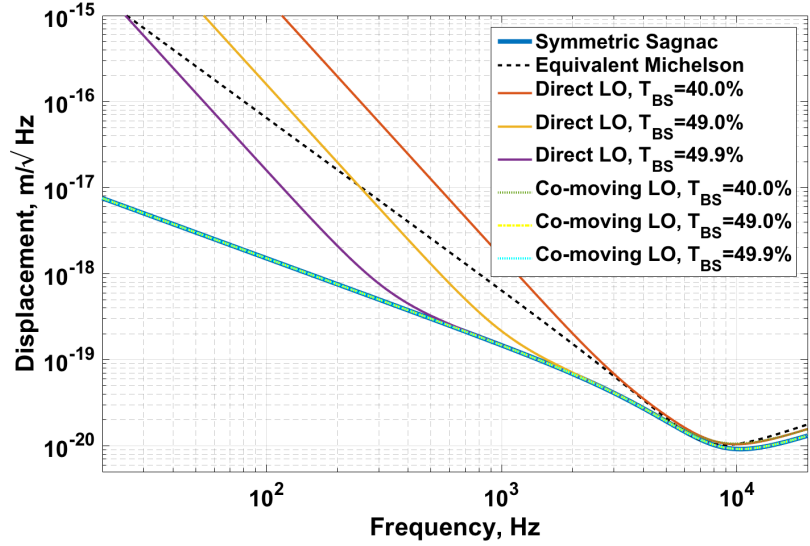}
	\caption{Plots of QNLS of Sagnac interferometer for two different options of local oscillator in balanced homodyne detector. Dashed black curve shows QNLS for an equivalent Michelson interferometer. The green, yellow and cyan dot curves which corresponds to Sagnac interferometer QNLS with $10\%, 1\%$ and $0.1\%$ main beam splitter imbalance are almost overlapped with the blue solid curve that corresponds to symmetric Sagnac interferometer QNLS. All parameters are given in Table.~\ref{table:ssm_design_parameters}}
	\label{fig:noise}
\end{figure}

\subsection{Balanced homodyne readout}

One sees that asymmetry of the BS couples a fraction of pump laser light to the dark port of the interferometer. This creates a non-zero DC component of the signal light (\textit{i.e.} a component at the carrier frequency) that can be easily obtained from the I/O-relations above if one sets $\Omega\to0$ and $\mathcal{K}_{\rm arm}\to0$,
\begin{equation}
\pmb{O} = (R_{\rm BS}-T_{\rm BS})\pmb{P}\,,
\end{equation}
where the corresponding DC fields are expressed in terms of pump field at the main BS, $\pmb{P}$.
Analogously, one can derive the DC component of the local oscillator (LO) beam for all three choices of the LO.
\begin{enumerate}[label=(\roman*)]
\item $\pmb{L}_{\rm dir}\propto \pmb{P}$ for the direct LO option;
\item $\pmb{L}_{\rm AR}\propto \pmb{B}^{\rm RE} \propto \sqrt{R_{\rm BS}}\pmb{P}$ for the BS AR coating reflection LO option;
\item $\pmb{L}_{\rm co}\propto \pmb{Q}\propto -2\sqrt{R_{\rm BS}T_{\rm BS}}\pmb{P}$ for the \textit{co-moving} LO option;
\end{enumerate}
As shown in \cite{2015_PhysRevD.92.072009_Steinlechner,PhysRevD.95.062001}, the fluctuation part of the readout photocurrent of the balanced homodyne detector is proportional to a sum of following terms:
\begin{equation}\label{eq:BHD_asym}
I_{\rm HD} \propto \pmb{\hat{o}}^{\dagger}\mathbb{H}\pmb{L}+\pmb{O}^{\rm T}
\mathbb{H}\pmb{\hat{l}}\,,
\end{equation}
where 
\begin{equation}\label{eq:IHD}
\mathbb{H}=\begin{bmatrix}
\cos\phi& -\sin \phi  \\
\sin \phi & \cos \phi
\end{bmatrix}\,,
\end{equation}
with $\phi$ defining the homodyne angle. And $\pmb{\hat{l}}$ stands for the  noise fields of the local oscillator. For $\phi=\pi/2$ (phase quadrature readout), the photocurrent can be further simplified as
\begin{equation}
I_{\rm HD}\propto|\pmb{L}|{\hat{o}}_{s}-|\pmb{O}|\hat{l}_s\,,
\end{equation}
The potential of noise cancellation can be readily seen from this expression, for the  phase noise in the two optical paths comes from the same source, \textit{i.e.} from the pump laser. Following we continue to demonstrate how the quantum noise cancellation is tailored by properly choosing the LO delivery port.
The $\pmb{\hat{l}}$ field for three choices of the LO we consider here can be written along the same lines as corresponding classical amplitudes of the LO $\pmb{L}$:
\begin{enumerate}[label=(\roman*)]
\item $\pmb{\hat{l}}_{\rm dir}\propto \hat{\pmb{p}}$ for the direct LO option;
\item $\pmb{\hat{l}}_{\rm AR}\propto \hat{\pmb{b}}^{\rm RE}$ for the BS AR coating reflection LO option;
\item $\pmb{\hat{l}}_{\rm co}\propto \hat{\pmb{q}}$ for the \textit{co-moving} LO option.
\end{enumerate}
At low frequencies, the main contribution to the quantum noise comes from the off-diagonal radiation pressure term in the transfer matrices, as $\mathcal{K}_{\rm arm}$ and $\mathcal{K}_{\rm asym}$ both rise steeply as $\Omega\to0$.
Indeed, we substitute the Eq.~\ref{eq:o}, \ref{eq:q} into Eq.~\ref{eq:BHD_asym}, leaving only the leading terms, one can get for the low-frequency contribution to the readout photocurrent from bright port (BP) for the \textit{co-moving} LO option the following expression:
\begin{equation}\label{eq:Ico}
I_{\rm co}^{\rm BP}\propto \mathcal{I}_{\rm co}[(4\mathcal{K}_{arm}-\mathcal{K}_{asym})\sin \phi-2\cos \phi]\hat{p}_c=\mathcal{I}_{\rm co}[\mathcal{K}_{sym} \sin \phi -2\cos \phi]\hat{p}_c\,,
\end{equation}
Similarly, for BSAR LO option one can get:
\begin{equation}\label{eq:IBSAR}
I_{\rm BSAR}^{\rm BP}\propto \mathcal{I}_{\rm BSAR}[\mathcal{K}_{sym}\sin \phi-2\cos \phi]\hat{p}_c\,,
\end{equation}
where
\begin{eqnarray}
\mathcal{I}_{\rm co}&=2\sqrt{R_{\rm BS}T_{\rm BS}}(R_{\rm BS}-T_{\rm BS})e^{2i\beta_{\rm sag}}|\pmb{P}|\,,\label{eq:commonco} \\
\mathcal{I}_{\rm BSAR}&=2\sqrt{R_{\rm BS}}T_{\rm BS}(T_{\rm BS}-R_{\rm BS})e^{2i\beta_{\rm sag}}|\pmb{P}|\,.\label{eq:commonBSAR}
\end{eqnarray} 
With homodyne angle $\phi=\pi/2$, we simply have
\begin{equation}\label{eq:IcoBSAR}
I^{\rm BP}\propto \mathcal{K}_{sym}\hat{p}_c\,,
\end{equation}
for both the \textit{co-moving} LO and the LO derived from the BSAR coating reflection. This expression shows partial cancellation of steep low-frequency dependence and only the \textit{speed-meter-like} term remains, which manifests in flat low-frequency dependence. This remaining term, as we discuss later, stems from the differential back-action force driven by the bright port amplitude fluctuations represented by a cosine quadrature operator $\hat p_c$. Even though, since this remaining term is proportional to $|R_{BS}-T_{BS}|$ which refers to the beam splitter asymmetry, as shown in Eq.~\ref{eq:commonco},~\ref{eq:commonBSAR}, its contribution is always much smaller than the quantum noise contribution from dark port in terms of any realistic beam splitter imbalance. However, for LO derived directly form main laser, the expression has no radiation pressure related contribution in the second term in Eq.~\ref{eq:BHD_asym}, hence
\begin{equation}\label{eq:Idir}
I^{\rm BP}_{\rm dir} \propto \mathcal{K}_{\rm arm}\hat p_c \,,
\end{equation}
and the contribution from the bright port-driven common motion of the interferometer mirrors remains uncompensated. 

The physics behind this cancellation stems from the very principle of the balanced homodyne readout, where any fluctuations and variations of light that drive both, the local oscillator and the signal light in the same way, are cancelled by design. 
Hence the partial cancellation of quantum noise that we demonstrated above comes from this insensitivity to the common phase signal produced by the common part of the radiation pressure force, created by the bright port fluctuations $\pmb{\hat{p}}$, \textit{i.e.} $\hat F^{\rm r.p.}_c = (\hat F^{\rm r.p.}_n+\hat F^{\rm r.p.}_e)/2$ where $\hat F^{\rm r.p.}_{e,n}$ stand for radiation pressure forces in each of the arms. The remaining uncompensated part stems from the non-zero differential radiation pressure force, $\hat F^{\rm r.p.}_d = (\hat F^{\rm r.p.}_n-\hat F^{\rm r.p.}_e)/2$, ensuing from the imbalance of the amplitudes of the reflected and transmitted light at the asymmetric main beam-splitter.

\subsection{Quantum noise limited sensitivity of Sagnac interferometer with bright port noise cancellation}

It is straightforward now to calculate the QNLS power spectral density expressions for all three choices of LO, using the derived earlier I/O-relations for both, the BP and the DP of the interferometer. It requires knowing the transfer matrices of the BHD photocurrent in all three considered schemes on the input fluctuation fields, $\pmb{\hat{i}}$ and $\pmb{\hat{p}}$.  In order to simplify the equation, the rotation matrix $\mathbb{H}$ is absorbed into $\pmb{L}$ and $\pmb{O}$. After expressing the LO fluctuations field, $\pmb{\hat{l}}$, in terms of $\pmb{\hat{i}}$ and $\pmb{\hat{p}}$ one gets from \eqref{eq:BHD_asym}:
\begin{align}
&\hat I_{\rm dir} \propto \pmb{L}_{\rm dir}^{\rm T}\mathbb{T}_i\pmb{\hat{i}} + \bigl(\pmb{L}_{\rm dir}^{\rm T}\mathbb{T}_p+\pmb{O}^{\rm T}\bigr)\pmb{\hat{p}} + t^{\rm dir}_d x_d + t^{\rm dir}_c x_c\,,\\
&\hat I_{\rm co} \propto \bigl(\pmb{L}_{\rm co}^{\rm T}\mathbb{T}_i+\pmb{O}^{\rm T}\mathbb{R}_i\bigr)\pmb{\hat{i}} + \bigl(\pmb{L}_{\rm co}^{\rm T}\mathbb{T}_p+\pmb{O}^{\rm T}\mathbb{R}_p\bigr)\pmb{\hat{p}} + t^{\rm co}_d x_d + t^{\rm co}_c x_c\,,\\
&\hat I_{\rm BS, AR} \propto \bigl(\pmb{L}^{\rm T}_{\rm AR}\mathbb{T}_i+\pmb{O}^{\rm T}\mathbb{T}_i^{\rm RE}\bigr)\pmb{\hat{i}} + \bigl(\pmb{L}_{\rm AR}^{\rm T}\mathbb{T}_p+\pmb{O}^{\rm T}\mathbb{T}^{\rm RE}_p\bigr)\pmb{\hat{p}} + t^{\rm AR}_d x_d + t^{\rm AR}_c x_c\,,
\end{align}
where the last two terms stand for the signal part of the BHD photocurrent caused by the differential and common signal motion of the mirrors, respectively.  For the general case of arbitrary homodyne angle, $\phi_{\rm LO}$, the corresponding expressions for the \textit{dARM} and \textit{cARM} responses in all three cases read:
\begin{align}\label{eq:BHD_Resp}
t^{\rm dir}_d &= ie^{i\beta_{\rm sag}}\dfrac{\sqrt{2\mathcal{K}_{\rm sym}}}{x_{\rm SQL}}\sin\phi_{\rm LO}\,, & t^{\rm dir}_{\rm c} &= e^{i\beta_{\rm sag}}(R_{\rm BS}-T_{\rm BS})\dfrac{\sqrt{2\mathcal{K}_{\rm asym}}}{x_{\rm SQL}}\sin\phi_{\rm LO}\,,\\ 
t^{\rm co}_d &= ie^{i\beta_{\rm sag}}\dfrac{\sqrt{8R_{\rm BS}T_{\rm BS}\mathcal{K}_{\rm sym}}}{x_{\rm SQL}}\sin\phi_{\rm LO}\,, & t^{\rm co}_{\rm c} &= 0\,,\\ 
t^{\rm AR}_d &= ie^{i\beta_{\rm sag}}\dfrac{\sqrt{8R_{\rm BS}(T_{\rm BS})^2\mathcal{K}_{\rm sym}}}{x_{\rm SQL}}\sin\phi_{\rm LO}\,, & t^{\rm AR}_{\rm c} &= 0\,.
\end{align}
Note that for the \textit{co-moving} LO and for the BS AR-coating reflected LO there is an additional advantage of zero sensitivity to the common motion of the arms (\textit{cARM} degree of freedom). It cuts off the potential coupling of noise from the much loosely controlled cARM degree of freedom into the readout channel of the Sagnac interferometer.  
Finally, one can calculate the QNLS power spectral density of a Sagnac interferometer, 
 in the units of differential displacement of the arms using the following well-known general formula:
\begin{equation}
S^x_{LO\ option} = \dfrac{\langle in|\hat I_{LO\ option}(\Omega)\circ \hat I_{LO\ option}(\Omega')|in\rangle}{|t^{LO\ option}|}\,.
\end{equation}
The general formula reads:
\begin{multline}
	S^x_{\rm co}(\Omega) = S^x_{\rm DP,\,co} + S^x_{\rm BP,\,co} + S^x_{\rm PO,\,co} \\=  
	\dfrac{\bigl(\pmb{L}^{\rm T}\mathbb{T}_i+\pmb{O}^{\rm T}\mathbb{R}_i\bigr)\mathbb{S}^i\bigl(\mathbb{T}^\dag_i\pmb{L}+\mathbb{R}_i^\dag\pmb{O}\bigr)}{|\pmb{L}^{\rm T}\pmb{t}_d|^2} +
	\dfrac{\bigl(\pmb{L}^{\rm T}\mathbb{T}_p+\pmb{O}^{\rm T}\mathbb{R}_p\bigr)\mathbb{S}^p\bigl(\mathbb{T}^\dag_p\pmb{L}+\mathbb{R}_p^\dag\pmb{O}\bigr) }{|\pmb{L}^{\rm T}\pmb{t}_d|^2} +   \dfrac{T_p\pmb{O}^{\rm T}\pmb{O}}{|\pmb{L}^{\rm T}\pmb{t}_d|^2}\,.
\end{multline}
where we assumed that the power reflectivity/transmissivity of the pick-off beam splitter is equal to $R_p/T_p$ and there is an additional noise term, $S^x_{\rm PO}$ due to vacuum fields, entering the open port of this beam splitter. Here $\mathbb{S}^a$ is the spectral density matrix for the input light $\hat{\mathbf{a}}(\Omega)$,
defined as
\begin{equation}
2\pi S^a_{ij}(\Omega) \delta(\Omega-\Omega^\prime) = \langle vac|{\hat{a}_i(\Omega) \circ \hat{a}_j^\dagger(\Omega^\prime)}|vac\rangle\,,
\end{equation}
where averaging goes over the vacuum quantum state of light $|vac\rangle$ and $\{i, j\} = \{c, s\}$.
Substitution of \eqref{eq:TrMat} and \eqref{eq:OMResp} gives for the components of the QNLS the following simple formulae:
\begin{align}
&S^x_{\rm DP,\,co} = \dfrac{x_{\rm SQL}^2}{2}\dfrac{1+\bigl[\mathcal{K}^*_{\rm sym}-(8R_{\rm BS}T_{\rm BS}-1)\cot\phi_{\rm LO}\bigr]^2}{\mathcal{K}^*_{\rm sym}} \,,\label{eq:QNLS_DP_co}
\\ 
&S^x_{\rm BP,\,co} = \dfrac{x_{\rm SQL}^2}{2}\dfrac{(R_{\rm BS}-T_{\rm BS})^2\bigl[\mathcal{K}_{\rm sym}-\cot\phi_{\rm LO}\bigr]^2}{\mathcal{K}_{\rm sym}}\,,
\label{eq:QNLS_BP_co}\\
&S^x_{\rm PO,\,co} = \dfrac{x_{\rm SQL}^2}{2}\dfrac{T_p}{R_p}\dfrac{(R_{\rm BS}-T_{\rm BS})^2}{\mathcal{K}^*_{\rm sym}\sin^2\phi_{\rm LO}}\,,\label{eq:QNLS_PO_co}
\end{align}
where $\mathcal{K}^*_{\rm sym} = 4R_{\rm BS}T_{\rm BS}\mathcal{K}_{\rm sym}$ is the new effective optomechanical coupling factor with account for BS asymmetry. 
The suppression of noise due to the double measurement scheme of the SSM and BHD, the \textit{speedmeter frequency dependence} of the quantum noise at low frequencies, is seen in Fig. \ref{fig:noise}. 
\begin{figure}[t]
	\centering
	\includegraphics[width=0.8\textwidth]{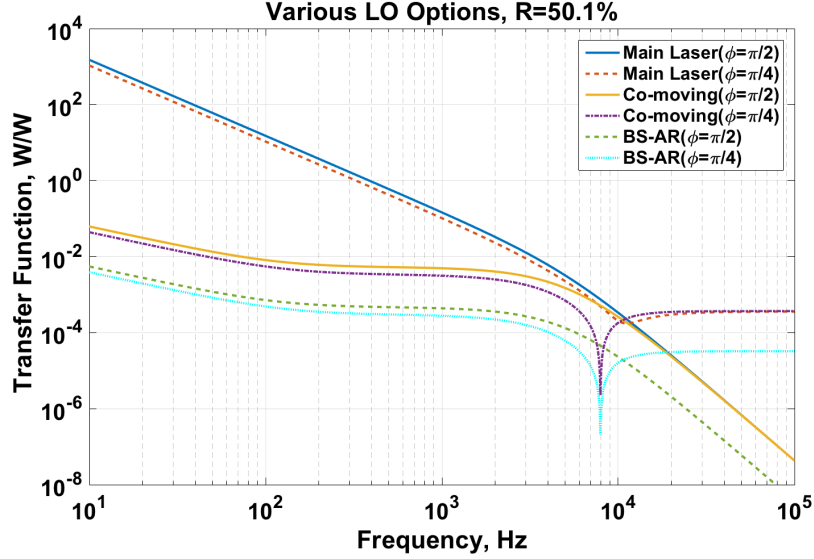}
	\caption{The laser amplitude fluctuations transfer function from the laser port to detection port for the three LO options with $0.1\%$ main beam splitter imbalance and different homodyne angle,\textit{i.e.} $\frac{\pi}{2},\frac{\pi}{4}$. The parameters are given in Table.~\ref{table:ssm_design_parameters} for Glasgow speed meter proof of concept  experiment..}
	\label{fig:TF}
\end{figure}

\begin{figure}[t]
	\centering
	\includegraphics[width=0.8\textwidth]{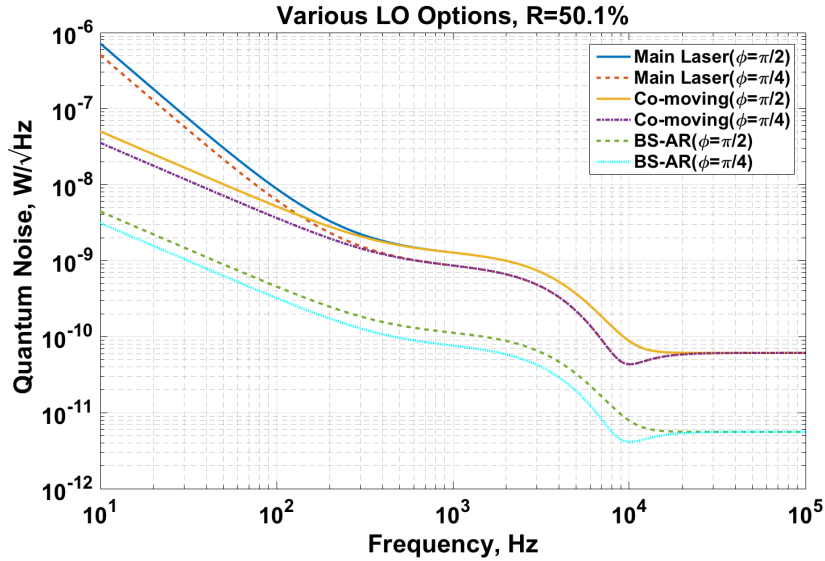}
	\caption{The quantum noise for the three LO options with $0.1\%$ main beam splitter imbalance and different homodyne angle,\textit{i.e.} $\frac{\pi}{2},\frac{\pi}{4}$. The parameters are given in Table.~\ref{table:ssm_design_parameters} for Glasgow speed meter proof of concept  experiment.}
		\label{fig:QN}
\end{figure}

\section{Relative laser intensity noise requirement }
\label{sec:glas}
\begin{table}[th]

\caption{\footnotesize\rm Parameters of 
the Glasgow SSM 
interferometer experiment. \label{table:ssm_design_parameters}}
\centering
\small
\begin{tabular}{p{4.5cm}|p{7.4cm}}
\hline
\textbf{Parameter} & \textbf{Value}\\
\hline\hline

Arm cavity length $L$ & $1.3$\,m\\

Optical power $P$ & 1.7\,W at beam splitter, $\sim$ 1\,kW in the arms\\

Arm cavity round trip loss & $\leq$ 25\,ppm\\

Optic mass $m$ & Arm cavity input test mass (ITM)  
$860$\,mg, arm cavity end test mass (ETM) $100$\,g\\

Transmissivities $T$ \& reflectivities $R$ & Central beamsplitter, $R_{\rm BS}=T_{\rm BS}=0.5$, ITM, $T_{ITM}=$700\,ppm\\

Main Laser LO \& \textit{co-moving} LO power  & 10mv\\

BSAR LO power & 0.078mv\\ 

\hline

Main readout & Balanced homodyne detector with suspended optical local oscillator path\\

 \hline
\end{tabular}
\end{table} 

The direct implication of suppression of laser noise contribution to the QNLS, discussed earlier and shown in Fig.~\ref{fig:noise}, is the much relaxed relative laser intensity noise (RIN) requirements, ensuing from the  significantly weakened transfer function from bright port amplitude quadrature to the BHD readout following from the Eqs.~\ref{eq:IcoBSAR} and \ref{eq:Idir}.

In this section, we consider as an example the Sagnac speed meter proof-of-principle experiment being built in the University of Glasgow \cite{Graef14}. Due to the complexity of the instrument, we have eschewed analytical calculation in favour of the numerical, using FINESSE \cite{finesse} to simulate the laser intensity noise RIN requirement.  This is done by simulating the quantum noise at the BHD detection port, finding the transfer function of input laser power noise at the bright port to detection port, and dividing quantum noise by the transfer function then by the input laser power. 

The transfer functions from the input laser amplitude fluctuations to the BHD readout port with homodyne angle $\pi/2$ and $\pi/4$ are shown in Fig.~\ref{fig:TF}. And the main beam-splitter asymmetry is characterised by setting  $R_{\rm BS}=0.501$.  As we can see, the transfer functions for \textit{co-moving} and BSAR LO options are significantly weakened compared to the main laser LO option in low frequency for both homodyne angles. Another feature that we notice is the difference between the two readout quadratures in high frequency for three LO options. That can be understood form the Eq.~\ref{eq:IcoBSAR}, since for phase quadrature readout, the transfer function of the amplitude quadrature noise is just proportional to $\mathcal{K}_{sym}$, which decrease along with the frequency growing in high frequency according to Eq.~\ref{eq:Ksym}. However, on an alternative homodyne angle as shown in Eq.~\ref{eq:Ico} and \ref{eq:IBSAR}, the amplitude noise gets coupled to the readout constantly and dominates in high frequency. From the two equations, we can also understand the dip at a specific frequency that indicates a cancellation between the frequency dependent back action noise and the constantly coupled amplitude noise for the case $\phi=\frac{\pi}{4}$.
We note that the gap between \textit{co-moving} LO option and BSAR LO option comes form the relatively weak LO power from BSAR as shown in Table.~\ref{table:ssm_design_parameters}. In this experimental set up, the power of the laser we use is only 1.7 W and the AR reflection is 100ppm. So that the presentation for BSAR option here is only on the state of principle illustration but not for realistic implementation for this experiment.

\begin{figure}[t]
	\centering
	\includegraphics[width=0.8\textwidth]{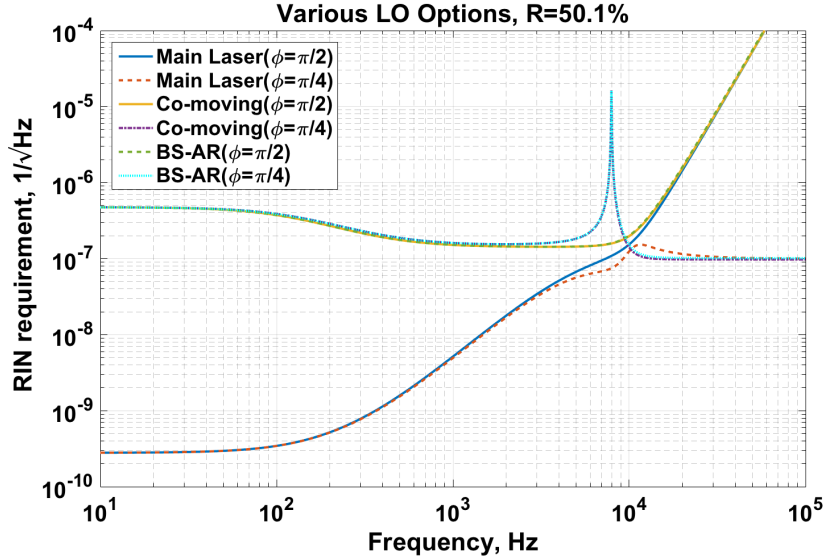}
	\caption{The RIN requirement of input laser for the three LO options with $0.1\%$ main beam splitter imbalance and different homodyne angle,\textit{i.e.}$\frac{\pi}{2},\frac{\pi}{4}$. The parameters used are given in Table.~\ref{table:ssm_design_parameters} for Glasgow speed meter proof of concept  experiment.}
	\label{fig:RIN}
\end{figure}

The Fig.~\ref{fig:QN} shows the quantum noise for the three LO options with different readout quadratures.
And Fig.~\ref{fig:RIN} shows the RIN requirement. As expected, the RIN requirement get relaxed by three orders of magnitude below 100 Hz by selecting co-moving or BSAR LO options.

\section{Summary}
\label{sec:summary}
Speed-meter configurations of GW interferometers are known to provide a significant improvement of quantum noise limited sensitivity at low frequencies because by suppression of quantum back-action noise using QND measurement of speed \cite{Chen2011,Review}. This advantage increases the signal-to-noise ratio (SNR) of speed-meter-based GW detectors for compact binary coalescences by at least two orders of magnitude if compared to the equivalent Michelson interferometer in the quantum-noise-limited case \cite{2018_LSA.7.accepted}. Zero-area Sagnac interferometer is one of the possible ways to realise the GWD based on speed-meter principle. However it was shown \cite{SagnacImperfections} that, in a non-ideal realistic case of asymmetric beam splitter, the fluctuations of the laser pump couple into the readout port of the interferometer, thereby creating an excess radiation pressure noise that significantly worsens the QNLS of speed meter interferometer and hence its SNR. In this work, we demonstrate that using a balanced homodyne readout scheme with a particular choice of the local oscillator option this detrimental effect can be almost completely attenuated. 

Picking the local oscillator beam from the reflected light at the pumping port of the interferometer (the \textit{co-moving} LO option), or from the direct reflection off the main beam splitter's AR coating (the BSAR LO option), one can significantly reduce the magnitude of the transfer function of the laser fluctuations from the pumping port to the readout one and qualitatively change its frequency dependence at low frequencies. We show analytically that this partial cancellation of laser fluctuations stems from the very nature of the BHD scheme that is inherently insensitive to any common variations of light phase in LO and signal beam of the BHD driven by input laser fluctuations. We further confirm our analytical findings by numerical simulation of the Glasgow proof-of-principle speed-meter interferometer set-up and estimating the relative laser intensity noise requirements for it. Our simulation shows that at frequency of 100Hz the RIN decreases by 3 orders of magnitude, form $4\times 10^{-10}/\sqrt{\rm Hz}$ to $4\times 10^{-7}/\sqrt{\rm Hz}$ if the \textit{co-moving} or BSAR LO option is chosen vs. the conventional \textit{direct} pick-off of the LO beam from the main laser. It is worth noting here that these 3 orders of magnitude of relaxation of the RIN requirement, mean reducing the RIN requirement from a very challenging value which is beyond the best achieved so far \cite{Seifert06,Kwee2011,Junker17} to a value which is easily achievable.

This feature of Sagnac interferometer can, in principle, be expanded to any scheme of speed-meter interferometer that uses the Sagnac-type way of performing the velocity measurement, where signal sidebands co-propagate with the carrier light throughout the main interferometer, including the polarisation-based speed meters \cite{Danilishin04,Wang13,2018_LSA.7.accepted}. Hence, we report here the method that solves the challenges originating from beam splitter asymmetry of a real speed-meters interferometer setup by using a balanced homodyne readout scheme with a particular choice of a local oscillator beam.

\section*{Acknowledhgements.} The authors are very grateful to our colleagues from the LIGO-Virgo Scientific Collaboration (LVC) for illuminating discussions and invaluable feedback on the research presented in this paper. TZ, SS, PD, JSH, EAH, SSL, SH and SLD were supported by the European Research Council (ERC- 2012-StG: 307245).  SLD was supported by the Lower Saxonian Ministry of Science and Culture within the frame of “Research Line” (Forschungslinie) QUANOMET – Quantum- and Nano-Metrology. The work of EK and FYK was supported by the Russian Foundation for Basic Research Grants 14-02-00399 and 16-52-10069. FYK was also supported by the LIGO NSF Grant PHY-1305863. SS was supported by the European Commission Horizon 2020 Marie-Skłodowska-Curie IF Actions, grant agreement 658366.

% \appendix
% \label{sec:appendix}
% \input{./sec_appendix}

\clearpage
\section*{References}
\bibliographystyle{unsrt}
\bibliography{power_stab}

\end{document}